\documentclass[journal]{IEEEtran}
\usepackage{array}
\usepackage[ruled,vlined]{algorithm2e}
\usepackage{caption}

\usepackage{booktabs}
\usepackage{amsmath,amssymb,amsfonts}
\usepackage{graphicx}
\usepackage{xspace}
\usepackage{textcomp}
\usepackage[nolist]{acronym}
\usepackage{color, colortbl}
\usepackage{adjustbox}
\usepackage[colorinlistoftodos]{todonotes}
\usepackage{xcolor}
\usepackage[bottom, multiple]{footmisc}
\usepackage{pgfplots}
\usepackage{pgfplotstable}
\usepackage{tikz}
\usepackage{subfigure}
\usepackage{multirow}

\usepackage{balance}
\usepackage{comment}
\PassOptionsToPackage{hyphens}{url}\usepackage{hyperref}

\def\BibTeX{{\rm B\kern-.05em{\sc i\kern-.025em b}\kern-.08em
    T\kern-.1667em\lower.7ex\hbox{E}\kern-.125emX}}
    
\usepackage{array}
\newcolumntype{L}[1]{
  >{\raggedright\let\newline\\\arraybackslash\hspace{0pt}}m{#1}}
\newcolumntype{C}[1]{
  >{\centering\let\newline\\\arraybackslash\hspace{0pt}}m{#1}}
\newcolumntype{R}[1]{
  >{\raggedleft\let\newline\\\arraybackslash\hspace{0pt}}m{#1}}

\pgfplotsset{compat=1.17} 

\begin{document}
\title{VINCY: A Smart-contract based Data Integrity and Validation
Tooling for Automated Vehicle Incident Investigation}

\author{André Budel, Reem Alhabib, Mark Nicholson,~\IEEEmembership{member,~IEEE} and Poonam Yadav, ~\IEEEmembership{senior member,~IEEE}
\thanks{All authors are with Computer Science Department, University of York, UK, contact e-mail: poonam.yadav@york.ac.uk .}}

\maketitle
\begin{abstract}
Automated Driving Systems (ADSs) are being manufactured at an accelerated rate, leading to improvements in traffic safety, reduced energy consumption, pollution, and congestion. ADS relies on various data streams from onboard sensors, external road infrastructure, and other vehicles to make driving decisions. For effective traffic accident reconstruction, investigators must produce, collect, store, and access real-time data. To ensure meaningful investigation, the data used by investigators must be accurate and maintain its integrity.

In this paper, we propose a smart-contract based data integrity and validation tool for automated vehicle incident investigation during road trials, considering uncertainties in a real-world environment.
\end{abstract}


%
\IEEEpeerreviewmaketitle

\section{Introduction}
Automated Driving Systems (ADSs) are being developed and integrated into vehicles at an increasing rate. These systems have the potential to significantly disrupt the fundamentals of road transport in the coming decade and beyond. As automation takes over the primary functions of dynamic driving tasks (DDT), the reliance on human drivers for decision-making decreases. This shift is expected to enhance road safety and reduce energy consumption, pollution, and traffic congestion. \\
However, ensuring public safety and security remains a crucial aspect when deploying vehicles with automated control technologies. While car manufacturers extensively test their technologies in controlled environments during development and verification stages, public road trials are necessary to validate the safety of newly developed ADSs in real-world conditions that involve unpredictable elements.\\
ADSs rely on various data feeds from onboard sensors, external road infrastructure, and other vehicles to make control decisions. To understand and analyze traffic events, investigators need access to real-time data that is produced, collected, stored, and made available. Reconstructing traffic events through real-time data can help identify causes and provide recommendations to minimize the risk of future incidents. For meaningful investigations, it is crucial that the data used by investigators is accurate and maintains its integrity, and investigators can trust its reliability. \\
This paper focuses on ensuring the integrity of incident data, and it explores the potential of blockchain technology to provide data immutability through cryptography and distributed peer-to-peer networks. Modern blockchains allow both data and programs (smart contracts) to be stored, eliminating the need for reliance on trusted third parties for data collection, storage, processing, and distribution. This is particularly important for ADSs, considering the significant intellectual property, reputational, security, and safety issues involved. There may even be liability and potential criminal proceedings implications for organizations developing and operating ADSs.\\
To address these challenges, the paper introduces a software tool prototype called VINCY, designed to assist trial organizations (TO) and safety investigators in incident investigations involving ADSs. The paper is structured as follows: Section \ref{Background} provides background information and discusses related work. Section \ref{Problem} presents the problem statement and motivation behind the project, along with a brief explanation of the current challenges. Section \ref{System} describes the system architecture and requirements for the proposed support tool. Section \ref{Eval} discusses the evaluation results, and Section \ref{Conclusion} concludes the paper by summarizing the findings and suggesting avenues for future work.

\section{Background and Related Work }\label{Background}

\subsection{Automated vehicles and Automated Driving Systems}
In this report the term “on-road” is used as defined by the
International Society of Automotive Engineers (SAE) in their
 Ground Vehicle Standard \cite{SAE}:
On-road refers to publicly accessible roadways
(including parking areas and private campuses that permit public access) that collectively serve all road users, including
cyclists, pedestrians, and users of vehicles with and without driving automation features.”
 It has been widely recognized that driver performance is a
key factor impacting the safety of on-road vehicle transport.
A 2015 report by the US National Motor Vehicle Crash
Causation estimated human error as the critical causation for
94\% of vehicle accidents\cite{singh2015critical}.\\
A U.S congressional report states that the three
motivations for driving motor innovations are:
\begin{itemize}
    \item technological advances enabled by new materials and
more powerful, compact electronics;
\item consumer demand for telecommunications connectivity
and new types of vehicle ownership and ridesharing; and
\item regulatory mandates pertaining to emissions, fuel
efficiency, and safety \cite{policy}.

\end{itemize}

As a result, the automotive industry has been introducing
vehicle features which assist the driver’s operation of a vehicle
with a vision toward more vehicle automation through ADSs
and a longer-term goal of a fully autonomous vehicle. New
vehicle technology has often translated into a safer
environment for road users \cite{policy}. Widely used augmentation
features among cars include vision such as cameras and
sensors, detection and response through collision warning and
assisted lane keeping safety technology.
More recently, ADSs are being developed and deployed
that go beyond augmenting the driver’s awareness and
operation, by performing driving functions automatically such
as emergency breaking to avoid collisions. A review carried
out by the National Highway Traffic Safety Administration
(NHTSA) in 2018 identified twenty-four unique conceptual
ADS features for on-road vehicles which were subsequently
grouped into seven generic categories \cite{thorn2018framework}.
As such, vehicles and their subsystems don’t fall into
binary categories “automated” or “not automated”. Therefore
in 2013, the NHTSA introduced a taxonomy of five levels of
vehicle automation. Conversely, in 2014, SAE introduced a
six-level taxonomy for driving automation systems \cite{SAE}. The
NHTSA has since adopted SAE’s six level taxonomy which
has become the widespread industry standard. The six levels
range from no driving automation (Level 0) through to full
driving automation (level 5). These levels are broadly defined
as:
\begin{itemize}

  \item  Level 0: No Driving Automation
  \item  Level 1: Driver Assistance
  \item  Level 2: Partial Driving Automation
  \item  Level 3: Conditional Driving Automation
  \item Level 4: High Driving Automation
   \item Level 5: Full Driving Automation
\end{itemize}
These standardised levels refer to driving automation
features that are engaged during on-road operation of the
vehicle at any given time. Therefore, although an ADS
equipped vehicle may be capable of achieving any given level
of driving automation, the level of driving automation at any
given time is classified by the features that are engaged in that
instance.


\subsection{Advanatges of Automated Vehicles}
Widespread use of AVs potentially brings a broad range
of benefits and influences to society including:
\begin{itemize}
    \item Accident Reduction: AVs have the potential to
reduce most accidents and mitigate the severity
of injuries by using technologies that out preform
human diver’s “perception, decision-making and
execution” \cite{taeihagh2019governing}. Bocca and Baek \cite{8804564} reviewed 50 accident reports from 2018 to analyse the impact careless human-driving had on the collisions with AVs and found that 92\% of collisions were caused by human drivers. Furthermore, the reported collisions caused by AVs vehicles caused only minor damage. 
\item Congestion: AV technologies may reduce road
congestion if large scale adoption takes place.
This reduction would be due to more efficient
routing of vehicles, AV sharing services, less
accidents and innovative combined vehicle
ownership models. However, its possible that this
benefit is diminished if individual ownership
models for vehicles remains the popular choice,
or there is an increase in vehicle usage if
passengers opt to be dropped off and have their
vehicle move to low-priced car parking or run
other errands \cite{lavieri2017modeling}

\item Economic Benefits: Andersson and Ivehammar
\cite{andersson2019benefits} used a cost-benefit analysis to determine the
generalised cost of operating an AV versus a
manually driven vehicle (MDV). They
concluded that operating an AV had a lower
generalised cost than an MDV. They illustrated
that for the capital costs to outweigh the
combined benefit from operating costs and travel
time saved by using ADVs, an increase in capital
costs of 22\% and 3\% was needed for cars and
trucks respectively. In addition, the study points
out there is a potential added economic gain to
the traveller if AVs operate at a level of full
driving automation as the traveller’s attention can
be saved for other uses. A further research report
\cite{America} estimated the societal cost of accidents in the
U.S. is around 1tillion USD annually. They then
argued that assuming AVs would only address
crashes resulting from “gross human error
(distraction alcohol and speeding etc)” there
would be a conservative annual reduction to
societal costs of 500 billion USD.
\item Improved access: Bocca and Baek \cite{8804564} report that Owner operated vehicles have a barrier for non-drivers including young, and
some elderly and physically challenged who are
forced to use drivers or public transport options.
A move to level-5 AVs will provide transport
independence for non-drivers.

\end{itemize}





\subsection{Implications of Automated Vehicles}
Although the introduction and adoption of AVs will likely
have its benefits, the move away from traditional driver
operated vehicles brings with it a range of challenges to
regulatory frameworks, liability laws, infrastructure and
vehicle safety, communication, security, and privacy
considerations.\\
Traditionally, governing frameworks have been setup to
regulate road transportation through laws that license drivers,
traffic regulation, liability, and insurance. As a result,
regulatory institutions are having to react by introducing new
policy and guidance at pace to keep instep with advancements
in technology to not only provide a level of safety for the
public but also a framework which allows the automotive
industry to innovate.\\
Liability for an on-road traffic accident is conventionally
assumed to be the driver of the vehicle at fault. Complexity is
added when determining liability if responsibility of decision
making and control input is being executed in part or in full
by the vehicle at fault \cite{taeihagh2019governing}. Liability assignment between
driver, prime vehicle manufacturers and third parties involved
in the design of the vehicles safety systems and sub systems
lacks clarity as currently much of the world has no legal
framework to assign liability from incidents involving ADS.
This leaves manufactures vulnerable to reputational risk from
accidents that involve ADS and insurers attempting to assign
liability for compensation and calculating risk-based
insurance premiums. UK law attempts to resolve this
ambiguity between liability and insurance for AVs under
certain accident scenarios, with Germany taking a similar
approach albeit with less clarity \cite{taeihagh2019governing}. Although the UK and Germany have provided some guidance, unlike most other countries, much of the liability is still assigned to the driver of a vehicle in the event of a crash which an ADS may have contributed to.

\subsection {Safety Investigation Considerations}
Modern systems entering the data centric age \cite{nicholson2016data} are becoming increasingly integrated with extensive sociotechnical components including software, machine learning, IoT, autonomy and distributed cognition. Advanced ADS fit into this category. The scope, complexity and scale of
automated systems continue to increase due to their varying
levels of interconnection, the data points that they produce and
the open context in which they operate in \cite{neurohr2021criticality}. Analyzing the
causal factors and the links between them for incidents
involving these systems is becoming progressively more
complex \cite{alexander2018state}. As a result, accident models are being
developed and used to investigate safety related incidents to
deal with data centric systems\cite{nicholson2016data}. By considering the data
from the entire system, including the organisation,
management, policy, equipment, interfaces and integrations,
analysis can be done to look at what went right as well as
what went wrong. Conducting a system-based analysis can
then be used to establish lessons learnt and subsequent
recommendations. Large data collections will need to be
analysed to enable accident investigation using systematic
models with graphical solutions to augment textual
techniques such as accident reports. Graphical solutions
provide a means to represent properties of events, such as
timing and sequences as factors.\\ 
Common techniques and models employed for causal
analysis include traditional models such as fault trees, event
tree analysis, Ishikawa diagrams and Petri-nets. More recent
examples of safety analysis models, which take a systems
safety approach, are Causal Analysis based on System Theory
(CAST) \cite{leveson2019learn} and the Functional Resonance Analysis Method
(FRAM)\cite{frost2014system}. The CAST approach to accident analysis takes
a layered systems approach through a series of practical steps
that investigate why existing preventative measures were not
successful. FRAM models and assesses the complex
functional interactions between elements within socio-
technical systems and is primarily used for accident
investigation \cite{frost2014system}. Although all these models are useful for
analysis and representation, there is no over-arching model
which fits the full set of causal factors, instead they are each
used to explore, analyze and represent the incident from different angles which can then feed into recommendations
or be used to inform proactive system-based analysis models
that are useful for identifying hazards and causes of accidents,
such as a System Theoretic Process Analysis (STPA) and a
System Hazard Analysis (SHA).
\subsection{Governance and Regulation}
As discussed, validating ADS operational safety on
public roads remains a challenge. While the use of emulators,
such as the Connected Autonomous Vehicle Emulator
(CASE)\cite{elmquist2017overview}, and private isolated road trials are useful when
gathering validation data for safety analysis on general
operation and edge case scenarios, they are unlikely to
provide the full level of safety assurance required for
commercial use on public roads \cite{koopman2018toward}. Therefore, on-road
trials of ADS are required to achieve safety assurance, at least
within current frameworks. Furthermore, Koopman and
Wagner \cite{koopman2018toward}, argue that to effectively and efficiently gather
sufficient datasets for safety validation, a nuanced approach
to data collection and safety analysis is required. Their
suggested approach involves using high fidelity testing
regimes to validate the assumptions and simplifications
resulting from low fidelity testing.\\
Governments around the globe are implementing legal
frameworks and guidance to allow and encourage on-road
trials of AVs.\\
For example, the UK government published a Code of
Practice in 2019 providing guidance to organizations wishing
to carry out on-road trials \cite{codeofpractice}. The Publicly Available
Specification (PAS):1881 \cite{PAS1882}, was published in 2020
compliments the code of practice. It provides a framework for
safety cases specifying the minimum requirements for AV
trials and development testing in the UK. The UK’s PAS
process “enables a specification to be rapidly developed in
order to fulfil an immediate need in industry”, displaying the
government’s commitment to efficiently provide a legal
framework and guidance to facilitate ADS trials.
In addition, PAS 1882 \cite{PAS1882}, released in 2021, specifies
data collection and management for AV trials for the purpose
of undertaking a safety incident investigation. This document
recognizes the role data collection has on the ability to
conduct meaningful causal investigations following an
incident. It specifies the information gathering requirements
that a trialing organisation (TO) needs to collect when
conducting trials on the public road. This system-based
approach to incident investigation is akin to that used in other
critical safety sectors such as aviation, rail and marine, and
contrasts with the current practice to vehicle investigation
which, in general, only focuses on the driver and the vehicle.
Not only does this document specify the information
requirements for data collections deemed essential, but it also
includes recommendations of additional information that
should be collected. As there are an infinite number of data
points which can support an investigation, planning needs to
be conducted before a trial to ensure that all data collections
are identified against known hazards and accident causes.
Further, the ability to identify currently unknown hazards and
causal factors must be considered.
The information gathering requirements address data sets,
access, storage, security, and stakeholders. Also included in
the PAS:1882 is additional information that is recommended
TOs collect to meet best practice. Moreover, it emphasizes
that to enable system-based, fact-finding incident
investigation, the provision of data collection, curation,
storage and sharing needs to be trustworthy. Such information
may also be requested and valuable to other stakeholders,
including criminal and civil investigators and insurers.
These information gathering requirements and
recommendations bring challenges to TOs and their
stakeholders. Firstly, in general, the technology used in ADSs
generates large quantities of data. Secondly, on-road trials
will likely require multiple information producing sources,
including data received, generated, or held but which is not
used for the direct operation of the vehicle, potentially making
retaining the data an expensive task. Additionally,
manufacturers and suppliers need to be able to retain privacy
over their IP and therefore releasing information can become
problematic. There may also be privacy concerns due to
personal data of operators that needs consideration. Lastly,
investigators need to have confidence that the data they are
analyzing is trustworthy, in that it hasn’t been manipulated,
censored, or corrupted. In response to these challenges, TOs
are required to develop an Information Management Plan
(IMP) which identifies what data will be collected and how it
will be managed so that the data required to derive the causes
of an incident during a trial is available to investigators.
Other notable guidance documents and specifications for
the operation of AVs in the UK include the PAS:1883 -
Operational Design Domain (ODD) taxonomy for an
automated driving system \cite{PAS1883}, and PAS:1884 - Safety operators in automated vehicle testing and
trialing \cite{PAS1884}.\\
Like the UK, the U.S. Department of Transport (US DOT)
has an approach to enable development and deployment of
ADS technology for public use. In 2017 the U.S DOT released
the Automated Vehicles Comprehensive Plan (AVCP)\cite{U.S.DeartementofTransportation}, as
their road map to facilitate an era of safe automated road
transport. The AVCP communicates three goals: “Promote
Collaboration and Transparency, Modernize the Regulatory
Environment, and Prepare the Transportation System”.
Furthermore, the US DOT have released a series of guidance
documents for Automated Driving Systems, AV 2.0 \cite{national2017automated}, AV
3.0 \cite{dot2018preparing} and AV 4.0 \cite{nstc2020ensuring} with each building on the previous.
AV2.0 is described as the “cornerstone voluntary guidance
document for Automated Driving Systems.” It takes a non-
regulatory approach by providing voluntary guidance for the
safe design and testing of AV technology. Within this
guidance, the NHSTA emphasize the importance that the
recording of data during trials has on understanding the safety
potential of ADSs and the lessons that can be learnt from a
thorough reconstruction of a crash. The NHTSA encourages
trialing entities engaged in on-road testing to establish a
documented process for “testing, validating, and collecting
necessary data”, a process which is comparable to the IMP
required by PAS:1882. Additionally, much like PAS:1882,
the guidance recommends that ADS data be collected, stored,
and made available for retrieval. At a minimum, entities
trialing ADS should have all information relevant to a crash
available to allow the events leading to the crash to be
reconstructed. Besides collecting, storing and maintain this
data, the guidance states that those entities should have the
ability, both technically and legally, to share all data required
for crash reconstruction with government authorities, again a
similar approach to that required by the PAS:1882.

\subsection{ On-Road ADS Trials}
With the introduction of laws, frameworks and guidance,
entities around the globe have been able to successfully
conduct various on-road trials of ADS technologies.
The first example of a live on-road trial in the UK is
currently underway in a government backed scheme (The UK
Autonomous Vehicle Scheme), Project Endeavour \cite{tyler2020project}. The
aim being “to accelerate and scale the adoption of AVs
services across the UK through advanced simulations
alongside trials on public roads.” The project involves a fleet
of vehicles equipped with level-4 ADS which are being
exposed to a range of traffic scenarios and weather and
lighting conditions, within predefined areas.
The first ADS is set to be approved in the UK to perform
a DDT instead of a human, is the Automated Lane Keeping
System (ALKS). The regulation enabling this is the ALKS
Regulation developed by the United Nations Economic
Commission for Europe which sets out the technical
requirements for ALKS. Certain aspects of its use still require
consideration. Once these aspects are addressed, the UK’s
Department of Transport expects amendments will be made to
the Highway Code allowing the incorporation and use of
ALKS of vehicles on the public road \cite{UKRules}. 

\subsection{Current Solutions}
The UK's Governments code of practice for automated vehicle trialling strongly recommends TOs "to develop plans for police investigators and relevant organisations to readily and immediately access data relating to an incident in a way that maintains the forensic integrity, security, and the preservation of the data" \cite{UKGovDoT}. 
With the aim of preventing fraudulent tampering and accurate auditing, blockchain technology has been utilised for forensic purposes, where AV data is stored on an externally storage that is accessible by authorised third parties.
In order to determine whether the sensors have been attacked in traffic incidents involving AVs Sharma et al. provide an AI-based forensic investigation protocol. It collects data from storage and memory devices, uses a supervised deep convolutional neural network model to analyse the accident, and finds anomalies in collected data. However, the useability of their technique have not been justified \cite{strandberg2022systematic}.
In \cite{feng2019autonomous} a non-intrusive mechanism is proposed for the collecting and storage of forensic data from AVs in smart cities through safely transferring it to cloud storage. Using a hybrid of centralised and decentralised databases and smart contracts, Parlak et al \cite{parlak2022tamper} suggested a blockchain-based insurance decentralised application to verify the authenticity and provenance of the accident footage and decentralise the failure-adjusting process. They aimed to avoid time and processing power consuming in in-line detection using AI analysis against deepfake. Thus they proposed a hybrid setup is advised, commencing with an application that records claim-related media and authenticates them at the point of recording while retaining a counter-deepfake tool for in-line detection. Another approach was created using blockchain technology where, in the event of accidents, the events are recorded for forensic purposes \cite{guo2018blockchain}. This plan is dependent on the endorsements from several authorities.  The Authors discussed just their theoretical analysis, but their consensus method for the blockchain enabled a reliable and verifiable proof-based system to track accidents by CAVs.
By combination of cooperative event correlation and trust model, a framework was suggested \cite{philip2023multisource} to detect malicious and false reporting, followed by Long Short Term Memory (LSTM) and Bayesian model to resolve conflicting event reports. Authors in \cite{guo2020proof} propose a forensic technique for an accident where the vehicles are using blockchain consensus mechanisms to come an to agreement on accident details. In this approach, data is collected from nearby vehicles through DSRC communication, hashed, and stored in a blockchain. They propose a quick leader election algorithm and Hyperledger Cloth Blockchain Network emulator test.  Na et al propose a multi-signature-based access control method by grouping and storing video data of multiple vehicles based on GPS (Global Positioning System) data \cite{9669940}. They used formulas for the GPS-based grouping method and access method to the Dashcam video data for decentralized Oracle configuration. \\

Communication among connected vehicles has become a significant requisite. Many researches recently have deployed blockchain technology to achieve steady and diminutive transmission. A blockchain-based protocol is presented in \cite{ahmed2021blockchain} to address security concerns, particularly to assure safe emergency VANET message transmission. One blockchain is used to store the vehicle's authentication data, and another is used to store and distribute blockchain services. Also for secure vehicle-to-vehicle (V2V) data transmission, the authors in \cite{9457110} utilise consortium blockchain technology to enable anonymous and traceable data sharing without the need of RSUs. Their proposed combination of 5G and blockchain technologies can  prevent malicious data sharing.
\\

Using blockchain-based anonymous voting, Ren et al \cite{9507046} proposed a feedback mechanism in IoVs that exploits the decision made by the transport manager to enhance the learning approach. To enable anonymous vote, they used the attributes of decision-related nodes instead of their identities. To validate car voting data in that work, the Blockchain layer communicates with the other microservices. To ensure the secure transfer of information among drones, a blockchain-based security mechanism for cyber-physical systems is proposed by Singh et al \cite{singh2020deep}. In their encryption architecture, a deep Boltzmann machine is used to select a miner node based on features such as computational capacity, available battery power, and drone flight time. In the same era, intelligent security solutions are required to detect new cyber threats automatically. Based on blockchain,  a framework has proposed in \cite{kumar2021privacy}   in order to guarantee privacy and security in Cooperative Intelligent Transport System C-ITS infrastructure. The suggested approach uses deep learning modules and the blockchain to provide two levels of security and privacy. First, a blockchain module is created to safely transfer C-ITS data between AVs, RSUs, and Traffic Command Centres TCCs, and an enhanced Proof of Work (ePoW) technique based on smart contracts is created to ensure data integrity and prevent data poisoning attacks. In order to prevent inference attacks, a deep learning module is created that uses the Long-Short Term Memory-AutoEncoder (LSTM-AE) technology to encode C-ITS data into a new format. The proposed Attention-based Recurrent Neural Network (A-RNN) uses the encoded data to identify intrusive events in the C-ITS infrastructure. Other work that aims to facilitate secure communication in vehicular networks is a proof-of-concept of PKI based on Ethereum certificate management scheme \cite{lin2020bcppa}. They suggest PKI based Elliptical Curve Digital Signature Algorithm (ECDSA) to be used by vehicles instead of private keys which further reduces verification time and cost

\section{Problem Statement} \label{Problem}

Data plays a crucial role in the success of incident
investigation of ADS systems by informing crash
reconstruction and system-based analysis modelling when determining causation. Government regulations require Trial Organisations (TOs) to collect, curate, store and share all the data with investigators that can be used to reconstruct an incident during an on-road ADS trial.  TOs are required to develop an Information Management Plan (IMP) which identifies what data will be collected and how it will be managed so that the data required to derive the causes of an incident during a trial is available to investigators. 
Modern on-road vehicles are often highly integrated with
sociotechnical components, which produce an increasing
number of data collection points, including data-intensive feeds such as video footage. Data sources for ADS trials are retrieved from the vehicle and its subsystems and an increasing number of external sources, such as road infrastructure, nearby vehicles, and meteorological information. Moreover, modern systems are often an assembly of subsystems and components from various manufacturers that are at risk, reputationally and legally, if found liable for causing an incident. Equipment manufacturers may also be concerned that information from a vehicle trial may contain their IP, which requires concealment. Safely curating and storing large quantities of trial data
from multiple sources while having it accessible and available for investigators creates a challenge for a TO. Furthermore, investigators need to be sure that the data they receive has retained its integrity from when it was collected, i.e., it has not been corrupted, manipulated, censored, or erased through malicious acts, equipment failure, error or otherwise.
This project endeavours to design and develop a proof-of-concept software tool that will provide a TO and its stakeholders with a collaborative means to safely curate, store and distribute ADS trial data while maintaining its integrity and privacy. 

\section{System Architecture and Development} \label{System}
An iterative development process\cite{light2009waterfall}, analogous to agile
the methodology was the approach taken to develop this project’s
proof-of-concept. An iterative process was favoured over an
an incremental approach, such as waterfall modelling \cite{light2009waterfall}, to
allow for flexibility throughout the design process.
High-level project requirements were established upfront
to inform the design and were used in each iterative design
stage to verify against. The project’s high-level requirements
and their respective justifications are provided in table 1. They
were established from the UK government framework and their
guidance to TOs for ADS trials. 
\begin{figure}[h]
  \centering
 \frame{ \includegraphics[scale=0.3]{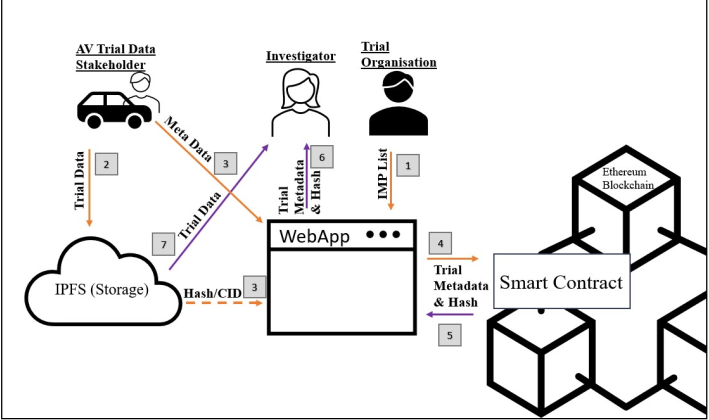}}
  \caption{System diagram for the proposed tool - VINCY.}
  \label{System Diagram}
\end{figure}


\begin{table*}[!htbp] 
\centering
\caption{High-Level System Requirements}
\label{tab: system}
\resizebox{\textwidth}{!}{%
\begin{tabular}{|l|l|}
\hline
\multicolumn{1}{|c|}{\textbf{Requirement}}                                                                                                                                                                                                             & \multicolumn{1}{c|}{\textbf{Justification}}                                         

\\ \hline
\begin{tabular}[c]{@{}l@{}} \textbf{Controlled Access} \\ The system shall ensure read and write access rights are enforceable.\end{tabular}                                    

& \begin{tabular}[c]{@{}l@{}}\\The PAS:1882 \cite{PAS1882} states that TOs shall have a plan which states the level \\ and type of information that each stakeholder can gain access to. \\ The proposed system shall be capable of enforcing this plan.\end{tabular}\\ \hline

\begin{tabular}[c]{@{}l@{}} \textbf{Data Security and Immutability} \\ The system shall by resilient against physical and cyber-attacks\\ and ensure the data is resistant to fraudulent acts, data manipulation,\\ deletion,and corruption.\end{tabular} 

& \begin{tabular}[c]{@{}l@{}}\\ \textbf The Code of Practice \cite{codeofpractice} states that “Each organization shall adopt the security \\principles set out in BS 10754-1. The TO shall secure all data collected and mitigate the security \\risks of the connected automotive ecosystem per PAS 11281.” \\Additionally, \cite{PAS1882} states, “TOs shall record any indicators of integrity to\\ give confidence that the data are authentic and have not been maliciously modified or corrupted.\end{tabular}        \\ \hline

\begin{tabular}[c]{@{}l@{}} \textbf{Scalability} \\ The system shall be capable of indefinitely  storing large \\amounts of data from multiple data sources.\end{tabular}                                                                                       & \begin{tabular}[c]{@{}l@{}}\\The PAS1882 specifies that the TO shall preserve all information,\\ that is, or might reasonably be considered relevant to an investigation of an incident.\\The TO needs to determine the period for securely retaining information\\ in-line with data protection laws. This period may vary between trials \\depending on stakeholder requirements such as insurers and law\\ enforcement agencies\end{tabular}                                                                                             \\ \hline

\begin{tabular}[c]{@{}l@{}} \textbf{Data Type Agnostic} \\ The system shall accept data in all-digital formats and sizes.\end{tabular}                                                                                                                          

& \begin{tabular}[c]{@{}l@{}}\\The PAS:1882 requires the TO to collect all relevant  information to the investigation of an \\incident. There is no exhaustive list of information types, therefore, the tool must\\ have the ability to manage any format. \\ Furthermore, the amount of data required will vary between trials and will likely increase \\as technology evolves, therefore, the tool must be able to scale to support increasing \\data requirements.\end{tabular}                                                          \\ \hline
\begin{tabular}[c]{@{}l@{}} \textbf{Accessible and Available} \\ The system shall remain available and accessible by all entities\\ requiring write and read access.\end{tabular}                                                                             & \begin{tabular}[c]{@{}l@{}}\\The PAS:1882, recommends “that the TO develops plans for police investigations and relevant \\ organizations to readily and immediately access data relating to \\ an incident in a way that maintains forensic integrity, security, and preservation of the data. \\ The TO shall prepare a data interface plan as part of the IMP, \\that sets out the approach to the sharing of information between the systems \\that make up an automated system and the organizations that operate them.\end{tabular} \\ \hline

\begin{tabular}[c]{@{}l@{}} \textbf{Auditability} \\ All Data entries and behaviour shall be traceable.\end{tabular}                                                                                                                                                & \begin{tabular}[c]{@{}l@{}}\\The PAS:1882 specifies that data which describes and/or identifies the “specific incidents shall \\be collected as a form of metadata and shall provide a way to create \\a trail of information. In addition, the PAS mentions that audit processes shall be \\put in place over the IMP and associated Information Management Systems.\end{tabular}                                                                                                                                                             \\ \hline
\begin{tabular}[c]{@{}l@{}} \textbf{Environmentally Considerate}\\ The system design should consider environmental impact when making \\design decisions.\end{tabular}                                                                                          & \begin{tabular}[c]{@{}l@{}}\\ In line with the UK government’s 25-year environmental plan, \\consideration needs to be given to environmental sustainability when designing \\and operating Information and Communication Technologies (ICT) systems.\end{tabular}                                                                                                                                                                                                                                        \\ \hline
\end{tabular}%
}
\end{table*}

\subsection{High-level System Overview}

Figure \ref{System Diagram} illustrates a high-level system diagram of the developed proof-of-concept solution. The two main components are the front-end web application and a smart contract deployed on a blockchain. The project makes use of an existing cloud storage solution.
In this system, the ADS trial data, outlined by the trial’s IMP, is uploaded into cloud storage. A unique hash of this data is created and uploaded to a blockchain, along with helpful metadata, using a web application as an interface. If an incident occurs, an investigator can retrieve the data to inform
their investigation from the cloud storage. To ensure the data has not been corrupted, manipulated, censored, or changed since being submitted, they can cross-check its hash against the original stored in the blockchain. In this way, data integrity can be assured. Additionally, the TO can store the list of required files, detailed in the IMP, for a trial on the blockchain. They can monitor what files have been submitted into the system by using the web application to query the blockchain. In this way,
the completeness of the IMP data set can be managed and ensured.
In the rest of the paper, we call this developed proof-of-concept - the Automated
\textbf{V}ehicle \textbf{INC}ident investigation data s\textbf{Y}stem (VINCY).
The term “transactions” when used in the context of
databases in this paper represents a set of changes to a
database which must be accepted or rejected. Table~\ref{tab: system} presents the high-level of the system requirements. 

\subsection{Blockchain and Smart Contracts}
Blockchain technology was first introduced in 2008 by an
anonymous entity, Satoshi Nakamoto, to instantiate the
decentralised digital currency known as Bitcoin \cite{nakamoto2008bitcoin}.
The basic idea is to form an immutable record of transactions
between peers on a digital ledger. If the ledger cannot be
changed, the integrity of the data on record is maintained with
no reliance on a trust-based model, which has inherent
weaknesses (e.g., vulnerability to cyber-attacks, corruption, and
deceit). Each peer that opts into the network can maintain a copy of the ledger, making its storage decentralised as not one
person or organisation owns or operates it. The integrity of the
transactions are maintained through cryptographic proof,
called Proof of Work (PoW) which makes them computationally impractical to reverse or modify. The process of nodes providing PoW is referred to as mining, an analogy to the process of mining resources such as gold, as nodes which provide computational power to PoW are rewarded
 with block rewards in Bitcoin. Using PoW as a consensus
mechanism provides security to the network enabling
participants to interact with the record through transactions without reliance on a trusted third party.
A further development was the introduction of scriptable
smart contracts, first introduced in the Ethereum Blockchain in 2015 \cite{wood2014ethereum}. A scriptable smart contract is an automation application capable of self-execution based on
predetermined conditions that operate on a blockchain.
Adding scriptable smart contract functionality turns a
blockchain into a globally decentralized computer that can
execute many applications at the same time. Developers can
create, manage, and update their smart contract using the
blockchain to synchronize and store its state changes while
inheriting the availability, auditability, transparency,
neutrality, security, and robustness of the blockchain. Smart
contracts built on Ethereum reside at an Ethereum address and
run functions when triggered by a transaction.\\
The VINCY makes use of the Ethereum blockchain by
developing a smart contract to validate the
the integrity of ADS trial data submitted and stored in the cloud
database before, during and after an incident. 
VINCY uses a public blockchain as a metadata file system
to a database that adds transactional assurance and traceability
for its readers. 
Although the blockchain is a database, it is only
practicable to store small quantities of data for applications as
each byte written to the network requires a financial payment,
so only the hash and essential metadata is being stored to
Ethereum by the VINCY.
The blockchain-based solution was chosen over the
alternatives for the combination of the following reasons:
\begin{itemize}
 \item  No  need for a central trusted intermediary.
 \item  Each accepted entry onto the blockchain is
unerasable.
 \item No capital purchase, licencing or subscription to
software is required.
\item The Ethereum blockchain have a near 100\% uptime, ensuring availability. 
 \item  Although in general a rational or noSQL database
will transact quicker, a transaction speed faster than
Ethereum offers will offer negligible difference to
the user of the interacting with the VINCY.
\end{itemize}
Solidity is Ethereum’s native Turing-complete
programming language that is used for implementing smart
contracts. It’s a curly-bracket language \cite{Solidity} designed to target
the Ethereum Virtual Machine (EVM)~\cite{zhou2018security}. 
Interacting with the components of the Ethereum
blockchain is performed through transactions. Web3.js gives
developers the methods for this interaction by facilitating a
connection to Ethereum nodes through a JSON RPC interface.
The smart contract has been developed to securely store
the metadata of each data collection submitted to a database
that can be retrieved later.
The four supported primary blockchain tasks are:
\begin{itemize}
  \item Allowing the IMP authority to store data
requirements from the IMP into states on the
blockchain.
  \item Allowing the owner of trial data records to record
the metadata associated with their trial data on
the blockchain.
  \item  Following an incident in a trial, an investigator
and IMP authority can retrieve the metadata
against each data collection associated with the
trial and can check which data collections
required by abn IMP have not been submitted.
  \item Allow an investigator to trace the history of
changes made to a data collection i.e., what was
the original data, when was it changed and by
who.
Therefore, it is important that traceability and
integrity of data are preserved throughout the
investigation process as incident data is produced
by analysis and fusion.
\end{itemize}
The following is a high-level overview of the proposed
contract’s mechanics:
\begin{itemize}
    \item The metadata submitted to the contract for each
data collection is held in a struct and pushed onto
a dynamic struct array.
  \item As each entry is pushed onto the array, a unique
ID is created which reflects the entry’s indexed
position in the array. This entry is added to a key-
value store with the key being the index and the
value being the trial ID.
\item To prevent unauthorized users from entering
bogus data into the contract, function modifiers
have been added to create exclusive contract
rights. The function modifier onlyOwner,
imported from the community Contract Ownable
\cite{Ownable.sol.}, restricts access to a whitelist of addresses so
that only the contract owner can amend it. The
contract owner is assigned when the contract is
deployed. This can only be reassigned later by the
current owner. The whitelist is a list of authorised
addresses. The whitelist is used as a function
modifier to the contracts POST functions.
\item Each time a struct is pushed onto the array, an
event is triggered which allows any users with a
listener setup to receive the event that a new entry
has been made, what it was and who submitted it.
\item In solidity, there is no way to return an array of
structs. Therefore, when a user wants to view the
collection of metadata entries associated with a
trial ID, the return is broken into two parts.
Firstly, the contract returns an array containing
the keys in the key-value store that have a value
of the requested trial ID. This informs the user of
the array index of each struct associated with trail
ID. The user can then use consecutive queries to
the contract requesting the struct of each index
they received previously.
\item An additional key-value map is used to record the
number of data collections submitted against a
Trial ID. A user can query the contract to receive
this information.
\end{itemize}
The following metadata is stored for each data collection on a
contract:
\begin{itemize}
    \item The Filename of the data collection for identification
of the dataset.
\item The Trial ID which the data collection belongs to.
\item The File Hash of the data collection used as a
checksum against the stored dataset.
\item The Timestamp of the block which contained the
transaction. Used to record the time of submission
for auditability.
\item The submitter of the transaction identified through of
the user’s public key, also for auditability.\\

\end{itemize}

Following the successful deployment and testing of the
contract on a local blockchain using Ganache, the smart
contract was deployed on the public Ethereum Test Network
“Ropsten” to further simulate a production-like environment
on Ethereum \cite{NetworkEthereumFoundation}. Ropsten was chosen to use over
the alternatives as it’s the closet like-for-like representation of
the Ethereum mainnet \cite{NetworkEthereumFoundation}.
To transact with the contract on the Ropsten network, test
Ether (rETH) is needed as gas for the transactions. rETH can
be obtained by either providing computation power to the
network through PoW mining or requesting some from a
faucet. For this project, rETH was obtained from a faucet \cite{RopstenEthereumFaucet}.

\subsection{IPFS}
The default file storage and distribution component of this
prototype’s concept design utilises the Inter-Planetary File
System (IPFS).

VINCY makes use of IPFS as its cloud database by using
an IPFS Cluster \cite{ProtocolLabs.}. An IPFS Cluster is an application that
works as a side network to the IPFS where a global pin set (the
list of files that the network is responsible for storing and
maintaining) is intelligently allocated amongst the cluster’s
peers \cite{ProtocolLabs.} to ensure data redundancy and availability is
retained. Using an IPFS cluster, the VINCY benefits from the
benefits of storing its data in a distributed network while
retaining a permission model. The permission set consists of
standard peers (VINCY administrators) who can change the
cluster pin-set, and follower peers (content providers and host
nodes) which provide, and store content as instructed but
cannot modify the pin set. In the case where the loss of
privacy of a dataset is detrimental, such as an equipment
manufacturer protecting their intellectual property, these files
will need encrypting before being submitted to the network
and the encryption key to  authorised entities offline.

\subsection{User Interface}

The front-end user interface component for the VINCY is a
decentralised web application (dApp). Its purpose is to
provide a GUI to the users which facilitates transactions to and
from the smart contract.

Using a web application makes it simple for users to use
as no software is required. 
React has been used as the frontend framework to create a
single page application for user interaction with the smart
contract. React is a JavaScript Library used to create websites
and was used because it implements a single page application
(SPA). Implementing a SPA creates a smooth use experience
as only a single initial HTML request is made to a server
which passes the user’s browser the data and information
required to operate the application without making further
server requests. All interactions with the smart contract on the
blockchain will have unavoidable delays while using the dApp
interface as the user will need to wait for block times when
submitting transactions and async request to nodes when
fetching data. Therefore, using the React framework helps
mitigate the latency in the user experience.
The objective of this project was to create a proof-of-
concept; therefore, limited aesthetic features were added to the
dApp. However, some CSS styling was applied to ensure the
page’s workflow was laid out logically for user trials and
presentation.

\begin{figure}[h]
  \centering
 \frame{ \includegraphics[scale=0.5]{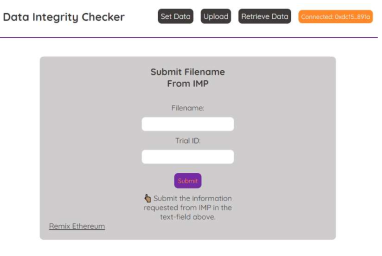}}
  \caption{dApp page used by the TO to submit the list containing
the names of datasets required by the IMP to the smart contract.}
  \label{dApp page}
\end{figure}

The dApp adhered to the Model-view-controller (MVC) software
design pattern \cite{krasner1988description}. A separate file in the dApp was used that
contained all its functions to manage its logic, data, and rules
which were then exported to the frontend.
To interact with the smart contract the dApp uses the
contract’s Application Binary Interface (ABI). Smart
contracts are stored on Ethereum as bytecodes in a binary
format and the ABI defines the functions you use to interact
with it as well as the format of return data.\\
Before any data is written to the contract, the user
submitting the data needs to sign the transaction using their
private key. This dApp uses Metamask, a virtual wallet in the
browser to manage the user’s Ethereum account which
provides the user with functionality to sign their transactions.
To use Metamask in the dApp a button is provided on the top
right corner, see figure \ref{dApp page}, which connects the users in the
browser to the dApp. If Metamask is not installed JSON object
is returned and a JSX object is relayed to the user prompting
them to install it before continuing.

\begin{figure}[h]
  \centering
 \frame{ \includegraphics[scale=0.5]{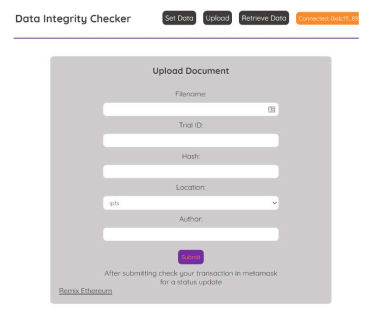}}
  \caption{dApp page used to view metadata stored on the blockchain}
 
  \label{dApp page used to view metadata stored on the blockchain}
\end{figure}

The WebSockets API key from Alchemy \cite{UsingWebSockets}  facilitates
the setup of a listener in the dApp that detects the events
emitted by the contract when the metadata has successfully
been stored. This data is then displayed in a JSX object to alert
the user that their submission was successful. Additionally, a
link is provided to the user once they’ve submitted a
transaction which directs them to Etherscan \cite{Etherscan}, an Ethereum block explorer which displays the block information related to
a transaction and would alert the user to any issue with their
transaction. Although this is suitable for a proof-of-concept, a
production ready tool would likely require a more obvious
error message in the GUI.
The first page of the dApp, as seen in figure \ref{dApp page}, is used by the TO to submit a list of dataset filenames documented in the IMP against a trial to be stored in the blockchain.
The second page of the dApp is used by stakeholders who
are submitting metadata associated with a trial’s data
collection stored on the IPFS to the smart contract, as seen in
figure \ref{dApp page used to view metadata stored on the blockchain}.

The final dApp page is used by investigators or TO. Using his page, the user can achieve two tasks: retrieve the metadata
of each data collection submitted to the contract against an
ADS trial or check to see what data collections are yet to be
submitted. The Alchemy Web3 API is used by the dApp to
make read calls to the contract to receive this data.

Figure \ref{The dApps page displaying the data retrieved from the blockchain} illustrates an example of a fictitious ADS trial
where the IMP has called for three datasets, however, only two have been
submitted to the VINCY.

\begin{figure}[h]
  \centering
 \frame{ \includegraphics[height= 6cm, width = 6.5cm]{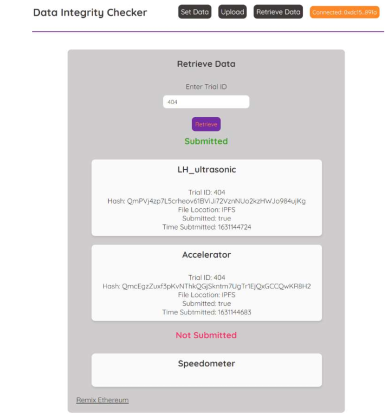}}
  \caption{The dApps page displaying the data retrieved from the
blockchain}
  
  \label{The dApps page displaying the data retrieved from the
blockchain}
\end{figure}

\section{EVALUATION AND DISCUSSION} \label{Eval}
User evaluations were carried out to evaluate the intuitiveness of the front-end dApp as well as testing for
software bugs, design oversights and efficiency of workflow.
The users selected for user testing had varying levels of
experience and knowledge with blockchain applications,
ranging from an active developer to users who had
only heard about bitcoin without directly interacting with it.
Five individual users were used for user testing, all were
independent from this project’s development up until they
carried out the user evaluation.
The users were supplied with an information sheet which provides a background
to the study as well as describing the objective of the dApp.
They were also supplied with a data pack which contained
“dummy” IMP data and access to an IPFS desktop node with
the files added that were required for the test. They were asked
to carry out three tasks on the dApp: firstly, to submit the
filenames of the data collections to the dApp required in the
IMP; secondly to update the metadata of stored collections
against given trials; and lastly, to retrieve the data
collections against a given trial, determine what had and
hadn’t been submitted against the trial’s IMP and whether
the data had retained its integrity.
The trials were conducted over the internet using
ADS trial or check to see what data collections are yet to be
discord \cite{discord} for video conferencing. The users accessed the
submitted. The Alchemy Web3 API is used by the dApp to
dApp, Meta mask and the provided IPFS node through Team
make read calls to the contract to receive this data.
Viewer \cite{team}. They were asked to “think out loud” as they
worked their way through the evaluation. Any perceived issues or difficulties the users had were noted down for discussion following each task. Once each task was
completed, the user was asked to elaborate on what they were
having difficulty with each recorded issue and asked to
rank it given three options: Catastrophic (task cannot be
completed without a change), Severe (task can be completed
but is difficult or doesn’t work properly), Cosmetic (the task was
completed without difficulty but was not intuitive or
something was misleading).

\begin{figure}[h]
  \centering
 \frame{ \includegraphics[height= 5cm, width = 6.5cm]{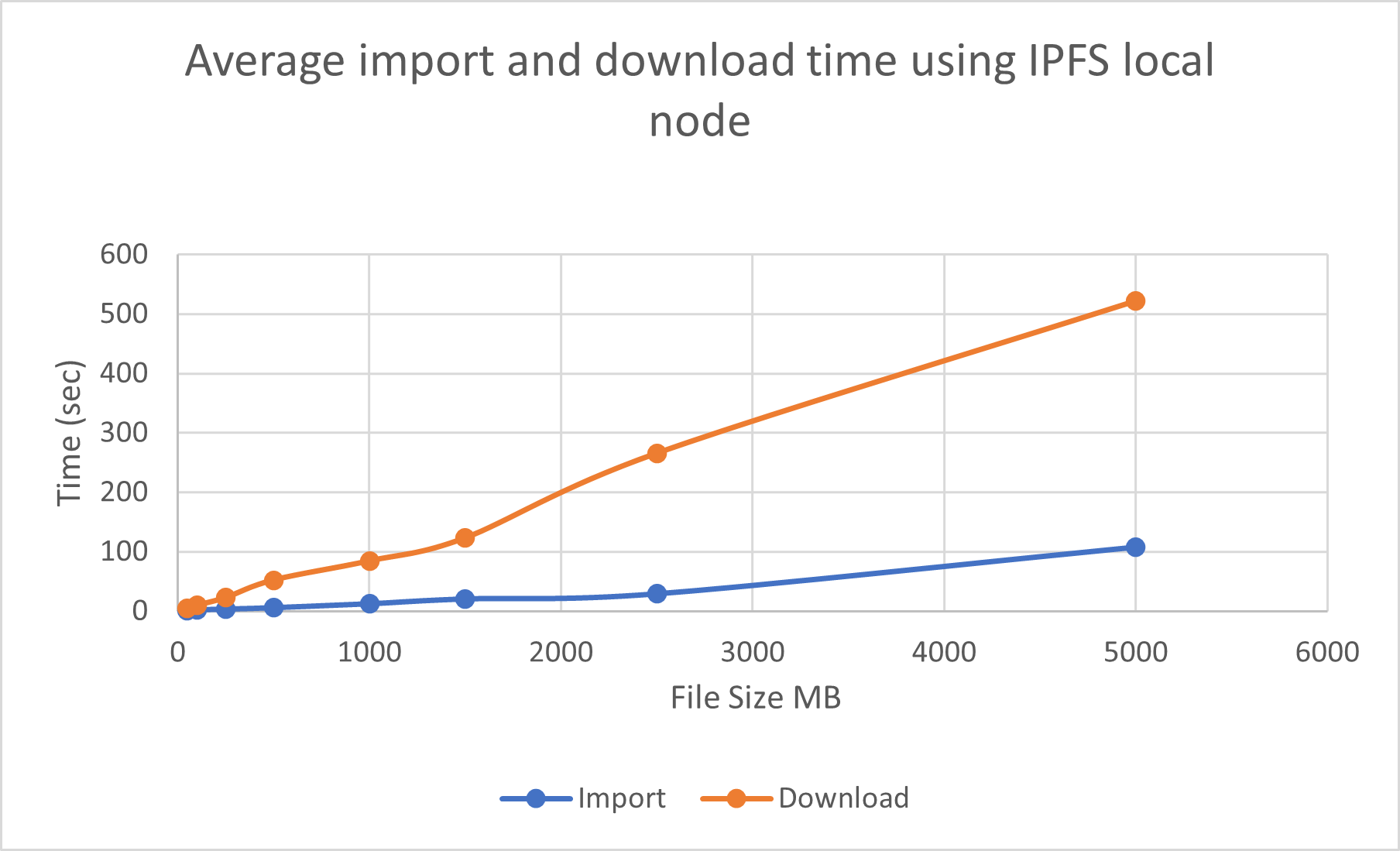}}
  \caption{Recorded import, hash and download times of IPFS Node}
  \label{EvalGraph}
\end{figure}

To evaluation the efficiency of the proposed system, the import and retrieve times were measured. Data files of varying sizes were generated and the average time to import the files to a local IPFS node and generate their associated were recorded. The average block time on the Ethereum network during 2022 was calculated at 13.18 secs using the data set available from Etherscan\cite{Etherscan} which is considered negligible when evaluating the upload time of large data sets against the IPFS. Additionally, the average download time to retrieve these files from IPFS were also recorded. Data retrieval from the Ethereum network is <  1 second which is also considered negligible. The results from this evaluation are displayed in Figure\ref{System Diagram}.

Following user evaluation, VINCY was compared against
its high-level requirements to verify the design and
implementation. The verification summary is detailed below:
\begin{itemize}
    \item  Controlled Access – Access control is achieved
by the white list encoded into the smart contract
giving write permissions. Private data can be kept
secure through encryption on the IPFS.
\item Data Security and Immutability – IPFS’s
distributed network provides mitigation against
files from being deleted. By storing the file’s data
hash on the blockchain it becomes immutable,
therefore the integrity of its corresponding
dataset can be checked and maintained.
\item Scalable – VINCY is capable of scaling both
vertically and horizontally improving efficiency
and throughput capacity as demand grows.
VINCY benefits from the data efficiency protocol
operated by IPFS which through the evaluation discussed proved to support efficient data import and recovery.
\item Data Type Agnostic – VINCY supports all digital
file formats through the IPFS.
\item Accessible and Available – As both IPFS and
Ethereum run on a distributed network they are
resilient against outages, ensuring it remains
available. By deploying the dApp on the IPFS
and using it as its webserver, it will also become
resilient against outages. As the VINCY interface
is a web application, it is accessible to all users
who have an internet connection and a web
browser without the need to install additional
software.
\item Auditable – Storing the trial data’s meta
maintains an auditable trail of the data during its
life cycle.
\item Environmentally Considerate – On demand
cloud storage solutions have been shown to be
more energy efficient than using private storage.
The Ethereum blockchain was used to host
VINCY’s smart contract because has moved to a
PoS consensus which addresses the power
consumption concerns of PoW.

\end{itemize}


\section{CONCLUSIONS AND FUTURE WORK}\label{Conclusion}

ADS on-road trials necessitate the collection, curation, storage, and distribution of a substantial amount of information to support incident investigations. To ensure the integrity of this information throughout its lifecycle and utilize it as factual evidence in investigations, VINCY, a proof-of-concept software tool, has been developed to aid Trial Organizations (TOs) in data management. VINCY combines the accessibility, efficiency, and reliability of a distributed cloud storage protocol with the immutability of blockchain to securely store data hashes and metadata. The web application interface allows users to interact with the blockchain application, enabling transactions and queries.

The high-level requirements of VINCY were derived from relevant ADS trial standards and guidance, and it has undergone verification against them. Future work will focus on the following recommendations:
\begin{enumerate}
\item Conducting user testing with TOs and stakeholders in a more representative simulation of a trial and shadowing a live trial. This will provide valuable insights for further system development and validation of requirements.
\item Simplifying the user experience by amending the decentralized application (dApp) so that users don't need to directly interact with the InterPlanetary File System (IPFS), streamlining the process.
\item Automating the submission of datasets and metadata to VINCY for data coming from a "black box," such as an onboard data recorder. This automation would enhance the integrity of the submission process, as it eliminates the need for human intervention.
\end{enumerate}
By addressing these recommendations, VINCY can be enhanced to better serve the needs of TOs and stakeholders involved in ADS trials, providing a more efficient and reliable platform for data management and investigation support.

\section{Acknowledgements}
Alhabib's research is funded by the Saudi Arabian Cultural Bureau and Shaqra University in Saudi Arabia, while Yadav is funded by EPSRC under grants EP/X040518/1 and EP/X525856/1.

\balance
\bibliographystyle{IEEEtran}
\bibliography{IEEEabrv,IEEEbib.bib}

\end{document}